\documentclass{ws-ijmpb}

\begin{document}

\markboth{J.~W.~Clark and H. Li}
{Application Of Support Vector Machines To 
      Global Prediction Of Nuclear Properties}

%
\catchline{}{}{}{}{}
%

\title{APPLICATION OF SUPPORT VECTOR MACHINES TO \\ 
      GLOBAL PREDICTION OF NUCLEAR PROPERTIES}

\author{John W. Clark, Haochen Li}

\address{Physics Department, Washington University,\\
St.~Louis, Missouri 63130, USA}

\maketitle

\begin{history}
\received{Day Month Year}
\revised{Day Month Year}
\end{history}

\begin{abstract}
Advances in statistical learning theory present the opportunity 
to develop statistical models of quantum many-body systems exhibiting 
remarkable predictive power.  The potential of such ``theory-thin''
approaches is illustrated with the application of Support Vector 
Machines (SVMs) to global prediction of nuclear properties as
functions of proton and neutron numbers $Z$ and $N$ across
the nuclidic chart.  Based on the principle of structural-risk 
minimization, SVMs learn from examples in the existing database 
of a given property $Y$, automatically and optimally identify a 
set of ``support vectors'' corresponding to representative nuclei 
in the training set, and approximate the mapping $(Z,N) \to Y$ 
in terms of these nuclei.  Results are reported for nuclear 
masses, beta-decay lifetimes, and spins/parities of nuclear 
ground states.  These results indicate that SVM models can match
or even surpass the predictive performance of the best conventional 
``theory-thick'' global models based on nuclear phenomenology.
\end{abstract}

\keywords{global nuclear modeling; machine learning; database mining.}

\section{Introduction}	
Consider a class of quantum many-body problems such that each
member of the class is characterized by $X_1$ 
particles of type 1 and $X_2$ particles of a different type 2,
and by the interactions, both pairwise and multiparticle,
that operate among any subset of particles.  With the interactions
in place, all of the states and all of the observables of any 
exemplar of the class are determined by the integers $X_1$ 
and $X_2$.  Specifically, for any observable $Y$ defined
for the class, there exists a physical mapping from $(X_1,X_2)$
to $Y$.  This notion is trivially extended to more than two
type of constituents.

Quite obviously, the nuclear many-body problem defines 
a class of this kind, with $X_1$ and $X_2$ taken as the numbers
$Z$ and $N$ of protons and neutrons in a nuclide.  Other 
examples coming easily to mind: $^3$He-$^4$He clusters,
binary and ternary alloys, etc.  

Approaches to calculation -- or prediction -- of the properties 
of individual systems belonging to such a class span a broad 
spectrum from pure {\it ab initio} microscopic treatments to 
phenomenological models having few or many adjustable parameters, 
with hybrid macroscopic/microscopic and density-functional methods
in between.  These approaches are ``theory-thick'' to varying 
degrees, with the {\it ab initio} ones based in principle on 
exact theory and the phenomenological ones invoking physical 
intuition, semi-classical pictures, and free parameters.  

Thinking in the spirit of Edwin Jaynes, inventor of the MaxEnt 
method and charismatic proponent of Bayesian probability,\cite{jaynes} 
it becomes of special interest to go all the way in the 
``theory-thin'' direction and ask the question:  
\begin{itemize}
\item[]
{\it To what extent does the existing data on property $Y$
across the members of a system class, {\it and only
the data}, determine the mapping $(X_1,X_2) \to Y$?}
\end{itemize}

In general, this mapping takes one of two forms, depending
on whether $Y$ is a continuous variable (e.g., the nuclear
mass excess or quadrupole moment)
or a discrete variable (e.g., the nuclear spin and parity).
The former case defines a problem of function
approximation, while the latter defines a classification
problem.  During the past three decades, powerful new methods have been 
developed for attacking such problems.  Chief among these are 
advanced techniques of statistical learning theory,
or ``machine learning,'' with artificial neural networks 
as a subclass.  Considering the concrete example of the
mapping $(Z,N) \to M$ that determines the nuclear (i.e.,
atomic) mass, a learning machine consists of (i) an input 
interface where $Z$ and $N$ are fed to the device in coded form, (ii) 
a system of intermediate processing elements, and (iii) an output 
interface where an estimate of the mass appears for decoding.  Given 
a body of training data to be used as exemplars of the desired
mapping (consisting of input ``patterns,'' also called vectors,
and their associated outputs), a suitable learning algorithm is used 
to adjust the parameters of the machine, e.g., the weights of the 
connections between the processing elements in the case of a neural 
network.  These parameters are adjusted in such a way that the learning 
machine (a) generates responses at the output interface that reproduce, 
or closely fit, the masses of the training examples, and 
(b) serves as a reliable predictor of the masses of test nuclei 
absent from the training set.  This second requirement is a strong 
one -- the system should not merely serve as a lookup table for 
masses of known nuclei; it should also perform well in the much 
more difficult task of prediction or {\it generalization}.

The most widely applied learning machine is the Multilayer 
Perceptron (MLP), consisting of a feedforward neural network with at 
least one layer of ``hidden neurons'' between input and output 
interfaces.\cite{haykin}  MLPs are usually trained by the 
backpropagation algorithm,\cite{rumel,hertz,haykin} essentially 
a gradient-descent procedure for adjusting weight parameters 
incrementally to improve performance on a set of training examples.
A significant measure of success has been achieved in constructing 
global models of nuclear properties based on such networks, with applications 
to atomic masses, neutron and proton separation energies, spins and 
parities of nuclear ground states, stability vs.\ instability, branching 
ratios for different decay modes, and beta-decay lifetimes.  (Reviews
and original references may be found in Ref.~5.)

The Support Vector Machine (SVM),\cite{vapnik1,vapnik2,haykin} 
a versatile and powerful approach to problems in classification 
and nonlinear regression, entered the picture in the 1990s.  
Rooted in the strategy of structural-risk minimization,\cite{haykin} 
it has become a standard tool in statistical modeling.  Although 
Multilayer Perceptrons as well as Support Vector Machines are 
in principle universal approximators, SVMs eliminate much of 
the guesswork of MLPs, since they incorporate an automatic process 
for determining the architecture of the learning machine.  Not 
surprisingly, they have become the method of choice for a wide variety
of problems.

Our selection of global nuclear systematics as a concrete
example for the application of advanced machine-learning
algorithms is neither accidental nor academic. 
There exists a large and growing body of excellent data 
on nuclear properties for thousands of nuclides,
providing the raw material for the construction of robust
and accurate statistical models.  Moreover, interest in 
this classic problem in nuclear physics has never been
greater.

The advent of radioactive ion-beam facilities, and the promise of the 
coming generation epitomized by the Rare Isotope Accelerator (RIA), 
have given new impetus to the quest for a unified, global 
understanding of the structure and behavior of nuclei across a 
greatly expanded nuclear chart.  The creation of 
hundreds of new nuclei far from stability opens exciting prospects for 
discovery of exotic phenomena, while presenting difficult challenges 
for nuclear theory.  Following the pattern indicated above, 
traditional methods for theoretical postdiction or prediction of the
properties of known or unknown nuclides include {\it ab initio} 
many-body calculations employing the most realistic nuclear 
Hamiltonians\cite{pieper} and, more commonly, density functional
approaches and semi-phenomenological models.  Since computational 
barriers limit {\it ab initio} treatment to light nuclei, viable 
global models inevitably contain parameters that are adjusted 
to fit experimental data on certain reference nuclei.  Global models 
currently representing the state of the art are hybrids of 
microscopic theory and phenomenology; most notably, they include the 
macroscopic/microscopic droplet models of M\"oller 
et al.\cite{moellernix,moelleretal} and the density functional 
theories employing Skyrme, Gogny, or relativistic mean-field 
Lagrangian parametrizations of self-consistent mean-field 
theory.\cite{pearson,bender,bertsch}  From the standpoint 
of data analysis, these approaches are inherently theory-thick, 
since their formulation rests on a deep knowledge of the 
problem domain.

It is evident that data-driven, ``theory-thin,'' statistical models 
built with machine-learning algorithms can never compete with traditional 
global models in providing new physical insights.  Nevertheless, 
in several respects they can be of considerable value in complementing 
the traditional methods, especially in the present climate of 
accelerated experimental and theoretical exploration of the nuclear 
landscape.  
\begin{itemize}
\item[(i)]
A number of studies\cite{clark1,NN,lusofission,athannp,adv1,adv2,licmt} 
suggest that the quality and quantity of the data has already reached 
a point at which the statistical models can approach and possibly
surpass the theory-thick models in sheer predictive performance.  
In this contribution, we shall present strong evidence from machine 
learning experiments with Support Vector Machines that this is 
indeed the case.
\item[(ii)]
In spite of their ``black-box'' origin, the machine-learning 
predictions can be directly useful to nuclear experimentalists 
working at radioactive ion-beam facilities, as well as astrophysicists 
developing refined models of nucleosynthesis.
\item[(iii)]
Although not straightforward, it will in fact be possible to
gain some insights into the inner workings of nuclear physics 
through statistical learning experiments, by applying techniques
analogous to gene knock-out studies in molecular biology. 
\item[(iv)]
It is fundamental interest to answer, for the field of nuclear
physics, the Jaynesian question that was posed above.
\end{itemize}

\section{Key Features of Support Vector Machines}

In technical jargon, the Support Vector Machine is a
{\sl kernel method},\cite{haykin} which, in terms of a classification
problem, means that it implicitly performs a nonlinear mapping 
of the input data into a higher-dimensional {\sl feature space} 
in which the problem becomes separable (at least approximately).  
Architecturally, the Support Vector Machine and Multilayer Perceptron
are close relatives; in fact the SVM includes the MLP with one
hidden layer as a special case. However, SVMs in general offer important 
advantages over MLPs, including avoidance of the curse of dimensionality 
through extraction of the feature space from the training set, once
the kernel and error function have been specified.

Support Vector Machines may be developed for function approximation
(i.e., nonlinear regression) as well as classification. 
In either case, the output of the machine (approximation to
the function or location of the decision hyperplane, respectively)
is expressed in terms of a representative subset of the 
examples of the mapping ${\bf x} \to y$ contained in 
the training set.  These special examples are the
{\sl support vectors}.  

The basic ideas underlying SVMs are most readily grasped by first 
considering the case of a classification problem involving 
linearly separable patterns.  Suppose some of the patterns
are green and the others are red, depending on some input
variables defining the $n$-dimensional input space.  To
find a decision surface that separates red from green patterns, one 
{\sl seeks the hyperplane that provides the maximum margin 
between the red and green examples}.  The training examples that 
define the margin are just the support vectors.  In this 
simple case, an exact solution is possible, but in general errors 
are unavoidable.  When faced with a problem involving nonseparable
patterns, the objective then is to locate a decision hyperplane 
such that the misclassification error, averaged over the training 
set, is minimized.  Guided by the principle of structural-risk 
minimization,\cite{haykin} the SVM approach determines
an optimal hyperplane by minimizing a cost function that includes 
a term to reduce the VC dimension\cite{vapnik1,vapnik2,haykin}  
(thereby enhancing generalization capability) and a term that 
governs the tradeoff between machine complexity and the number 
of nonseparable patterns.  

In practice, the SVM strategy actually involves two steps.
The first is to implement a nonlinear mapping {\boldmath $\varphi$}: 
${\bf x} \to \varphi_j({\bf x}),\, j=1,\ldots,m>n$, from the 
space of input vectors ${\bf x}$ into a higher-dimensional feature 
space, which is ``hidden'' from input and output (and corresponds 
to the hidden layer in MLPs).  This is done in terms of 
an inner-product kernel $K({\bf x},{\bf x}_k) 
=${\boldmath $\varphi$}$^{T}(\bf x)${\boldmath $\varphi$}$({\bf x}_k)$ 
satisfying certain mathematical conditions, notably Mercer's 
theorem.\cite{mercer}  The second step is to find a hyperplane that 
separates (approximately, in general) the features identified in 
the first step.  This is accomplished by the optimization
procedure sketched above.

A self-contained introduction to the SVM technique is beyond
the scope of the present contribution.  Excellent treatments are
available in the original work of Vapnik as expounded in 
Refs.~6,7 and in Haykin's text\cite{haykin} (see also Ref.~19).

To provide some essential background, let us consider 
a regression problem corresponding to a map ${\bf x} \to y({\bf x})$, 
where ${\bf x}$ is an input vector with $n$ components $x^{(i)}$, 
and suppose that $T$ training examples indexed by $k$ are made available.
Then the optimal approximating function takes the form
\begin{equation}
{\hat y}_{\rm opt}({\bf x})
= \sum_{k=1}^{T} (\alpha_k - \alpha_k') K({\bf x},{\bf x}_k) \,,
\label{est}
\end{equation}
Solution of the optimization problem stated above determines the 
parameters $\alpha_k$ and $\alpha_k'$, and the support vectors 
of the machine are defined by those training patterns for 
which $\alpha_k \neq \alpha_k^\prime$.
 
Different choices of the inner-product kernel appearing in
Eq.~(\ref{est}) yield different versions of the Support
Vector Machine.  Common choices include
\begin{equation}
K({\bf x},{\bf x}_k) =
({\bf x}^T{\bf x}_k + 1)^p \,,  
\label{polynomial}
\end{equation}
corresponding to the {\sl polynomial learning machine} with
user-selected power $p$; a Gaussian form
\begin{equation}
K({\bf x},{\bf x}_k) 
=\exp \left( - \gamma ||{\bf x} -{\bf x}_k ||^2\right) 
\label{rbf}
\end{equation}
containing a user-selected width parameter $\gamma$, which generates
a radial-basis-function (RBF) network; and 
\begin{equation}
K({\bf x},{\bf x}_k) =\tanh (\beta_1 {\bf x}^T{\bf x}_k + \beta_2) \,,  
\label{mlp}
\end{equation}
which realizes a two-layer (one-hidden-layer) perceptron,
only one of the parameters $\beta_1$, $\beta_2$ being 
independently adjustable.  We also draw attention to a generalization
of the RBF kernel (\ref{rbf}) introduced recently as
a simplified version of what is called ANOVA 
decomposition,\cite{stitson} 
having the form
\begin{equation}
K({\bf x},{\bf x}_k) 
= \left(\sum_{i=1}^n \exp \left[ - \gamma \left( x_k^{(i)}
 - x^{(i)} \right)^2 \right] \right)^d \,,
\label{anova}
\end{equation}

The Support Vector Machine may be considered as a feedforward
neural network in which the inner-product kernel, through 
an appropriate set of $m$ elements $K({\bf x},{\bf x}_k)$, 
defines a layer of hidden units that embody the mapping from
the $n$-dimensional input space to the $m$-dimensional feature 
space.  These hidden units process the input patterns nonlinearly 
and provide outputs that are weighted linearly and summed by 
an output unit.  As already pointed out, the familiar structures of 
radial-basis-function networks and two-layer perceptrons can 
be recaptured as special cases by particular choices of kernel.
However, the SVM methodology transcends these limiting cases in
a very important way: it automatically determines the number 
of hidden units suitable for the problem at hand, whatever the 
choice of kernel, by finding an optimally representative set of support 
vectors and therewith the dimension of the feature space.  In 
essence, the Support Vector Machine offers a generic and principled 
way to control model complexity. By contrast, approaches to 
supervised learning based on MLPs trained by backpropagation or 
conjugate-gradient algorithms depend heavily on rules of thumb, 
heuristics, and trial and error in arriving at a network 
architecture that achieves a good compromise between complexity 
(ability to fit) and flexibility (ability to generalize).

\section{Application to Nuclear Systematics}

In this section we summarize the findings of recent
explorations of the potential of Support Vector Machines
for global statistical modeling of nuclear properties.
The discussion will focus on the predictive reliability
of SVM models relative to that of traditional ``theory-thick''
models.  The properties that are directly modeled in these
initial studies, all referring to nuclear ground states, 
are (i) the nuclear mass excess $\Delta M = M-A$, where 
$M$ is the atomic mass, measured in amu, (ii) $\beta$-decay 
lifetimes of nuclides that decay 100\% via the 
$\beta^-$ mode, and (iii) nuclear spins and parities.

The requisite experimental data are taken from the on-line 
repository of the Brookhaven National Nuclear Data Center (NNDC) at
http://www.nndc.bnl.gov/.  The experimental mass values
are those of the AME03 compilation of Audi et al.\cite{audi}

Extensive preliminary studies have been performed to identify
inner-product kernels well suited to global nuclear modeling.
Earlier work converged on the ANOVA kernel (5) as a favorable
choice, and corresponding results have been published
in Ref.~19.  More recently, we have introduced
a new kernel that yields superior results, formed by
the sum of polynomial and ANOVA kernels and named the
pA kernel.  (Satisfaction of Mercer's theorem is conserved 
under summation.)  The new kernel contains three parameters
($p$, $\gamma$, and $d$) that may be adjusted by the user.  
Aside from parameters contained in the inner-product
kernel, the SVM procedure involves a constant $C$
giving the user control over the tradeoff between complexity
and flexibility, plus an additional control constant $\varepsilon$
in the regression case, measuring the tolerance permitted
in the reproduction of training data.  Thus, SVM models developed 
with the pA kernel contain four or five adjustable parameters 
(five in all applications reported here).

To allow for a meaningful evaluation of predictive performance 
(whether interpolation or extrapolation), the existing database
for the property being modeled is divided into three subsets, 
namely the {\sl training set}, {\sl validation set}, and {\sl test 
set}.  These sets are created by random sampling, consistently 
with approximate realization of chosen numerical proportions among 
them, e.g. (100--2$R$):$R$:$R$ for training, validation, and test sets, 
respectively, with $R<25$.  The training set is used to find 
the support vectors and construct the machine for given values of 
the adjustable parameters $p$, $\gamma$, $d$, $C$, and $\varepsilon$.  
The validation set is used to guide the optimal determination 
of these parameters, seeking good performance
on both the training and validation examples.  The test set 
remains untouched during this process of model development;
accordingly, the overall error (or error rate) of the final model 
on the members of the test set may be taken as a valid measure
of predictive performance.  When one considers how SVM models
might be applied in nuclear data analysis during the ongoing
exploration of the nuclear landscape, it seems reasonable that
consistent predictive performance for 80:10:10 or 90:5:5 
partitions into training, validation, and test sets would be sufficient
for the SVM approach to be useful in practice.

The SVM approach has been applied to generate a variety of global
models of nuclear mass excess, beta-decay lifetimes, and spin/parity, 
corresponding to different kernels, databases, partitions into 
training/validation/test sets, and tradeoffs between the relative 
performance on these three sets.  Here we will focus on those 
models considered to be the best achieved to date.  Moreover, 
due to limited space, we will restrict the discussion to the most 
salient features of those models and to an assessment of their 
quality relative to favored traditional global models and to 
the best available MLP models.  Further, more detailed information
may be found at the web site http://abacus.wustl.edu/Clark/svmpp.php, 
which generates SVM estimates of the listed nuclear properties for
$(Z,N)$ pairs entered by visitors.  This web site will be 
periodically updated as improved SVM models are developed.

\subsection{SVM Models of Atomic Mass Surfaces}

Development and testing of the SVM mass models to be highlighted
here are based on the AME03 data for all nuclides 
with $Z,\,N > 16$ and experimental masses having
error bars below 4\%.  This set of nuclides is divided into the 
four classes: even-$Z$-even-$N$ (EE) even-$Z$-odd-$N$ (EO), 
odd-$Z$-even-$N$ (OE), and odd-$Z$-odd-$N$ (OO).  
Separate SVM regression models were constructed for each 
such ``even-oddness'' class.   This does introduce some minimal 
knowledge about the problem domain into the modeling process; 
one might therefore say that the models developed
are not absolutely theory-free.  However, the data itself
gives strong evidence for the existence of different mass
surfaces depending on whether $Z$ and $N$ are even or odd.
Knowledge of the integral character of $Z$ and $N$ may,
quite properly, bias the SVM toward inclusion of associated 
quantum effects.\cite{gazula}

Table 1 displays performance measures for models based on
an 80:10:10 target partitioning of the full data set among
training, validation, and test sets, respectively.  Inspection
of the actual distributions of these sets in the $Z-N$ plane
shows that substantial fractions of the validation and
test sets lie on the outer fringes of the known nuclei, 
significantly distant from the line of stable nuclides.  
Accordingly, performance on the test set measures the 
capability of the models in extrapolation as well as 
interpolation.  Performance
on a given data set is quantified by the corresponding 
root-mean-square (rms) error $\sigma$ of model results relative 
to experiment (as is standard in global
mass modeling).  The ``optimized'' model parameters are included
in Table 1.  They show enough differences from one even-oddness
class to another to justify development of separate models
for the four classes.  

\begin{table}[ht]
\tbl{Performance of SVM global models of atomic mass.  All rms $\sigma$ 
values are in MeV, and \# stands for the number of nuclides in the 
indicated set. Training, validation, and tests sets are selected randomly 
with target proportions 80:10:10, respectively. For all 531 nuclides 
in EE class, the SVM error $\sigma$ is 0.42 MeV, compared to 1.5 MeV 
obtained by Bertsch et al.$^{13}$}
{\begin{tabular}{crlrlrlrlrcl} \toprule 
{~}&\multicolumn{2}{c}{Training}&\multicolumn{2}{c}{Validation}&\multicolumn{2}{c}{Test}&\multicolumn{2}{c}{Control Param's}&\multicolumn{3}{c}{pA Param's}\\
Classes & \#{ } & { } $\sigma$ & \#  &{ } $\sigma$ &\# &{ } $\sigma$ 
& $C(10^{-2})$&$\varepsilon(10^{-4})$ & $p$&  $d$ &{  } $\gamma$  \\ \colrule
EE & 425  & 0.26 & 53  & 0.72 & 53 & 0.83 & 1.307 & 9.976 & 2 & 6 & 25.84 \\
EO & 398  & 0.48 & 50  & 0.35 & 50 & 0.53 & 2.240 & 9.953 & 2 & 5 & 19.98 \\
OE & 397  & 0.19 & 50  & 0.31 & 50 & 0.35 & 1.298 & 9.801 & 2 & 6 & 25.93 \\
OO & 377  & 0.41 & 47  & 0.77 & 47 & 0.96 & 2.238 & 9.160 & 2 & 5 & 18.06 \\ 
Overall&1597& 0.35 & 200 & 0.58 & 200 & 0.71 &{}&{}&{}&{}&{} \\ \botrule  
\end{tabular}}
\end{table}

The results in Table 1 attest to a quality of performance, in
both fitting and prediction, that is on a par with the best 
available from traditional modeling\cite{moelleretal,pearson} and 
from MLP models trained by an enhanced backpropagation 
algorithm.\cite{athannp,adv2,athannp2}  To emphasize this 
point qualitatively, we display in Table 2 some representative rms 
error figures that have been achieved in recent work with all three 
approaches.  (We must note, however, that the data sets used for the 
different entries in the table may not be directly comparable, 
and the division into training, validation, and test sets does 
not necessarily have the strict meaning assigned here.)  The second 
SVM model listed in the table was developed for a partitioning
of the data into training, validation, and test sets of approximately
90:5:5, obtained by random transfer of nuclides from the 
validation and test sets of the 80:10:10 model to the 
training set.  

\begin{table}[ht]
\tbl{Performance measures of superior global models of the atomic mass
table developed through traditional approaches of nuclear theory (FRDM
and HFB2), enhanced backpropagation training of neural networks (MLP), 
and SVM regression methodology. 
All $\sigma$ values are in MeV and \# is the number of nuclides. 
SVM results for two different partitions of the full data set into
training, validation, and test sets are shown.}
{\begin{tabular}{ccccccc} \toprule
{~}&\multicolumn{2}{c}{Training Set}&\multicolumn{2}{c}{Validation
Set}&\multicolumn{2}{c}{Test Set}\\
& \# &  $\sigma$ & \#  & $\sigma$ &\# & $\sigma$  \\ \colrule
FRDM \cite{moelleretal} & 1303  & 0.68 & 351  & 0.71 & 158 & 0.70  \\
HFB2 \cite{pearson} & 1303  & 0.67 & 351  & 0.68 & 158 & 0.73  \\
MLP \cite{athannp} & 1303  & 0.44 & 351  & 0.44 & 158 & 0.95  \\
MLP \cite{adv2,athannp2} & 1303 & 0.28 & 351 & 0.40 & 158 & 0.71 \\ 
SVM (80:10:10) & 1597  & 0.35 & 200  & 0.58 & 200 & 0.71  \\
SVM (90:5:5)& 1797& 0.31 & 100 & 0.29 & 100 & 0.49 \\ \botrule
\end{tabular}}
\end{table}

The quality of representation that can be realized through
the SVM methodology may be highlighted in another way.
Employing the nuclear mass excess values generated 
by the SVM models of Table 1, we have calculated the 
$Q_\alpha$ values for eight alpha-decay 
chains of the superheavy elements 110, 111, 112, 114, 115, 116, 
and 118.  (The alpha-decay $Q$-value is defined as $Q_\alpha(A,Z) = 
M({\rm parent}) - M({\rm daughter}) - M(\alpha) = 
{\rm BE}(A-4,Z-2) + {\rm BE}(4,2) -{\rm BE}(A,Z)$,
where BE stands for the binding energy of the indicated nuclide.)  
Results are presented in graphical and tabular form on the 
web site http://haochen.wustl.edu/svm/svmpp.php.  For the
models of Table 1 (based on an 80:10:10 partition of the 
assumed AME03 data set), the average rms error of the
38 estimates of $Q_\alpha$ is 0.82 MeV, while the average
absolute error is 0.64 Mev.   We emphasize that these
estimates are predictions (rather than fits), since none
of the nuclei involved belongs to the validation or
test set.  Moreover, due to the situation of these
superheavy nuclides in the $Z-N$ plane, prediction of
the associated $Q_\alpha$ values provides a strong test
of extrapolation.

The performance of SVM mass models documented in Tables 1
and 2 and in the alpha-chain predictions gives assurance
that this approach to global modeling will be useful
in guiding further exploration of the nuclear landscape.
However, it is important to gain some sense of when and
how it begins to fail.  The performance figures for the
two sets of SVM models involved in Table 2 are consistent
with the natural expectation that if one depletes the validation 
and test sets of the 80:10:10 partition in favor of an
enlarged test set, the predictive ability of the model
is enhanced.  Conversely, one should be able to ``break''
the SVM modeling approach by random depletion of the
training set of the 80:10:10 model in favor of larger
validation and test sets.  Eventually the training set will 
become too small for the method to work at all.  The results of
a quantitative study of this process are shown in Figure 1.
Writing the generic partition as (100--2$R$):$R$:$R$, the error
measure increases roughly linearly with $R$ for $R$ greater 
than 10.

\begin{figure}[ht]
\centerline{\psfig{figure=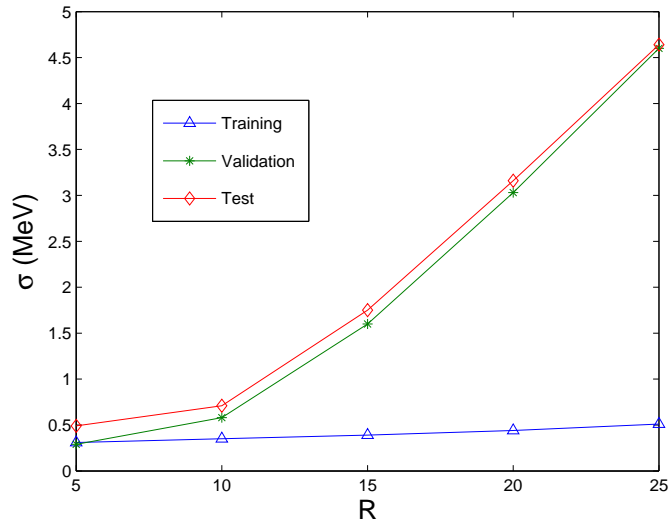,width=10cm}}
\vspace*{8pt}
\caption{Overall performance of EE, EO, OE, and EE SVM models
as a function of the target percentage $R$ of nuclides assigned 
to both validation and test sets.}

\end{figure}

In addition to direct statistical modeling using either SVMs or
MLPs, a promising hybrid approach is being explored.  Recently,
the {\it differences} $\Delta M_{\rm exp} - \Delta M_{FRDM}$ between
experimentally measured masses and the corresponding theoretical 
masses given by the Finite-Range Droplet Model (FRDM) of M\"oller, 
Nix, and collaborators\cite{moellernix,moelleretal} have been modeled
with a feedforward neural network of architecture 4--6--6--6--1
trained by a modified backpropagation learning algorithm.\cite{adv2}  
(The integers denote the numbers of neurons in successive layers, from
input to output.)  The rms errors on training (1276), validation
(344), and test (529) sets are respectively 0.40, 0.49, and 0.41
MeV, where the numbers of nuclides in each of these sets is given 
in parentheses.  In a similar experiment, we have constructed SVM 
models for EE, EO, OE, and EE classes using pA kernels.  
Overall rms errors of 0.19, 0.26, and 0.34 were achieved
on the training (1712), validation (213), and test (213) sets,
respectively, with little variation over even-oddness classes.
Error figures over comparable subsets for the FRDM model in question 
run around 0.7 MeV, again with relatively little 
variation from subset to subset.  These results suggest that MLPs and 
SVMs are capable of capturing some 1/2 to 2/3 of the physical
regularities missed by the FRDM.  It remains to be seen whether
the residual error has a systematic component or instead reflects
a large number of small effects that will continue to elude 
global description.

\subsection{SVM models of the Systematics of Beta Halflives}

Another important problem in global modeling involves the
prediction of beta-decay halflives of nuclei.   As in the case of
atomic masses, this is a problem in nonlinear regression.  Here 
we restrict attention to nuclear ground states and to nuclides that
decay 100\% through the $\beta^-$ mode.  For this presentation,
we make the further restriction to nuclides with halflives $T_{1/2}$
below $10^6$ s (although we have also included the longer-lived
examples in another set of modeling experiments).  The Brookhaven 
NNDC provides 838 examples fitting these criteria.  Since the examples 
still span 9 orders of magnitude in $T_{1/2}$, it is natural to
work with $L = \log T_{1/2}({\rm ms})$ and seek an approximation to the
mapping $(Z,N) \to L$ in the form of SVMs.  Again, we construct
separate SVMs for the EE, EO, OE, and OO classes, and again
a kernel of type pA is adopted.  The full data set is divided 
by random distribution into training, validation, and test sets 
in approximately the proportions 80:10:10.

The performance of the favored models is quantified in Table 3.  
Here we use two measures to assess the accuracy of the SVM results
for training, validation, and test nuclides.  These are the
rms error, again denoted by $\sigma$, and the mean absolute error 
$\mu$ of the model estimates of $L$, relative to experiment.

Detailed studies of beta-decay systematics within the established
framework of nuclear theory and phenomenology include those
of Staudt et al.\cite{staudt} (1990), Hirsch et al.\cite{hirsch}
(1993), Homma et al.\cite{homma} (1996), and M\"oller 
et al.\cite{kratz} (1997).  However,
comparison of the performance of the SVM models with that of the
models resulting from these studies is obscured by the
differences in the data sets involved.  Most significantly,
the data set employed here is considerably larger than those
used previously, including as it does many new
nuclides far from stability.  Analysis of SVM performance on 
subsets of the data set, now in progress, will yield useful 
information on the efficacy of SVM models relative to the more 
traditional ones, as we continue to develop improved global 
models of beta-decay systematics. 

On the other hand, MLP models for the beta-halflife problem
have been generated for the same data set as used in our
SVM study, allowing a meaningful comparison to be made.
The best MLP models created to date\cite{costiris} show
values for the rms error $\sigma$ over all even-oddness classes
of 0.55 in training, 0.61 in validation, and 0.64 in prediction. 
These values are somewhat larger than those seen in Table 3.  
However, it must be pointed out that the MLP results were 
obtained with a smaller training set, so the efficacy of the 
two statistical methods appears to be about equal at this 
stage of development.  That being the case, it is relevant
to note that the recent MLP models represent a distinct 
advance earlier versions,\cite{lusofission} 
and that those earlier statistical models already showed
better performance over short-lived data sets than the
conventional models of Homma et al.\cite{homma} and 
M\"oller et al.\cite{kratz}

\begin{table}[ht]
\tbl{Performance of SVM global models constructed for $\beta$-decay halflives 
below $10^{6}$s. Control parameter $C=3.861\times10^{-3}$ and pA parameters $p=1$, $d=8$ for all four classes.}
{\begin{tabular}{crccrccrcccc} \toprule 
{~}&\multicolumn{3}{c}{Training Set}&\multicolumn{3}{c}{Validation Set}&\multicolumn{3}{c}{Test Set}& \multicolumn{2}{c}{Param's}\\
Classes & \# &  $\sigma$ & $\mu$ & \#  & $\sigma$ & $\mu$ &\# & $\sigma$
& $\mu$ & $\varepsilon$($10^{-4}$)& $\gamma$ \\ \colrule
EE & 131  & 0.55 & 0.18 & 16  & 0.57 & 0.24 & 16 & 0.62 & 0.32 &9.155 &102 \\
EO & 179  & 0.41 & 0.15 & 22  & 0.42 & 0.19 & 22 & 0.51 & 0.25 &8.401 &79 \\
OE & 172  & 0.41 & 0.15 & 21  & 0.47 & 0.20 & 21 & 0.47 & 0.26 &7.862 &97 \\
OO & 190  & 0.52 & 0.18 & 24  & 0.40 & 0.16 & 24 & 0.52 & 0.28 &8.896 &102 \\
Overall&672& 0.47 & 0.16 & 83 & 0.46 & 0.19 & 83 & 0.53 & 0.27 &       &   
\\ \botrule  
\end{tabular}}
\end{table}

\subsection{SVM Models of Ground-State Parity and Spin}

The applications to prediction of atomic masses and beta-decay 
lifetimes demonstrate the predictive power of SVMs in two 
important problems of global nuclear modeling that involve 
function estimation.  The final two applications will probe 
the performance of SVMs in global modeling of the discrete 
nuclear properties of parity and spin.  In essence, these are 
problems of classification: ``Which of a finite number of 
exclusive possibilities is associated with or implied by a 
given input pattern?''  Support Vector 
Machines were first developed to solve classification problems,
and good SVM classifier software is available on the web.\cite{joachims}
However, for convenience and uniformity we prefer to treat
the parity and spin problems with the same SVM regression technique 
as in the other examples, also using the pA choice of inner-product 
kernel.  In the parity problem, the decision of the regression SVM is 
interpreted to be positive parity [negative parity]
if the machine's output is positive [negative].  In the spin problem,
the spin assigned by the machine is taken to be correct if and only
if the numerical output (after rescaling) is within $\pm 0.25$ of 
the correct value (in $\hbar$ units).

As before, all data are taken from the Brookhaven site.  For parity
and spin, it is especially natural to create separate SVM models
for the different even-oddness classes.  However, as is well known,
all EE nuclei have spin/parity $J^\pi = 0^+$.  Modeling this property 
is trivial for SVMs, so the EE class may be removed
from further consideration.  The data in and of itself permits us 
to do so.  Moreover, in the case of spin, the data itself establishes, 
with a high degree of certainty, that the spin of EO or OE nuclides 
takes half-odd integral values (in units of $\hbar$), while the 
spin of OO nuclides is integral.  Although this formulation of the
parity and spin problems introduces significant domain knowledge into the 
model-building process, the data alone provides adequate motivation.
Nuclei with spin values larger than 23/2 were not considered.
The predictive performance that may be achieved with SVM models
of parity and spin is illustrated in Tables 4 and 5.  Performance
is measured by the percentage of correct assignments.  Construction 
of both parity and spin models is based on an 80:10:10 partition 
of the data into training, validation, and test sets.  (As usual, 
the target distribution is realized only approximately.)  

Averaged over even-oddness classes, the overall performance of 
the parity SVMs is 97\% correct on the training set and 95\% on 
the validation set, with a predictive
performance on the test set of 94\%.
Obviously, assigning parity to nuclear ground states is an 
extremely easy task for Support Vector Machines.  One might expect quite 
a different situation for the spin problem: since there are 12 legitimate
spin assignments for the EO or OE nuclides considered (i.e., obeying
the rules for addition of angular momenta) and also 12 for 
the OO class, the chance probability of a correct guess is low.  It 
is then most remarkable that the SVM spin models we have developed 
perform with very high accuracy in prediction as well as fitting and 
validation.  While some success has been achieved previously in
MLP modeling of parity and spin,\cite{minn,gernoth} consistent 
predictive quality within the 80--90 percentile range has been 
elusive.  Within main-stream nuclear theory and phenomenology, 
the problem of global modeling of ground-state spins has 
received little attention, and the few attempts have not been very
successful.  As a baseline, global nuclear structure calculations 
within the macroscopic/microscopic approach\cite{moellernixspin} 
reproduce the ground-state spins of odd-$A$ nuclei with an 
accuracy of 60\% (agreement being found in 428 examples out 
of 713).

It should be mentioned that in the preliminary investigations
described in Ref.~19, the tasks of global modeling of parity 
and spin with SVMs were in fact treated as classification rather
than function-estimation problems.  Corresponding SVM classifiers 
were created using established procedures.\cite{haykin,joachims}
Based on an RBF kernel, results were obtained that surpass
the available MLP models in quality, but are inferior to
those reported here in Tables 4 and 5. 

\begin{table}[ht]
\tbl{Performance of SVM global models of nuclear ground-state parity. pA parameter $p=1$ for all three classes.} 
{\begin{tabular}{crlrlrlrlcc} \toprule 
{~}&\multicolumn{2}{c}{Training}&\multicolumn{2}{c}{Validation}&\multicolumn{2}{c}{Test}&\multicolumn{2}{c}{Control Param's}&\multicolumn{2}{c}{pA Param's}\\
Classes & \# &  Score & \#  & Score &\# & Score  & $C$ ($10^{-6}$) &$\varepsilon$ ($10^{-4}$) & $d$ & $\gamma$ \\ \colrule
EO & 468  & 99\% & 58  & 98\% & 58 & 97\% &7.629 &9.876  & 17 &19.57 \\
OE & 462  & 94\% & 56  & 96\% & 56 & 93\% &7.629 &9.921  & 17 &19.41 \\
OO & 429  &96\%  & 53  & 94\% & 53 & 92\% &1.907 &9.652  & 19 &19.45 \\
Overall &1357& 97\% & 168 & 95\% & 168 & 94\% & & &  &     \\ \botrule  
\end{tabular}}
\end{table}

\begin{table}[ht]
\tbl{Performance of SVM global models of nuclear ground-state spin. pA parameter $p=1$ for all three classes.}
{\begin{tabular}{crlrlrlrlcc} \toprule 
{~}&\multicolumn{2}{c}{Training}&\multicolumn{2}{c}{Validation}&\multicolumn{2}{c}{Test}&\multicolumn{2}{c}{Control Param's}&\multicolumn{2}{c}{pA Param's}\\
Classes & \# &  Score & \#  & Score &\# & Score  & $C$ ($10^{-3}$)
&$\varepsilon$ ($10^{-4}$) & $d$ & $\gamma$ \\ \colrule
EO & 523  & 93\% & 58  & 90\% & 58 & 86\% &7.634 &9.675  & 7 &202 \\
OE & 460  & 92\% & 57  & 88\% & 57 & 84\% &7.634 &8.804  & 7 &239 \\
OO & 469  & 85\% & 52  & 88\% & 52 & 81\% &7.634 &9.820  & 7 &192 \\
Overall&1452& 90\% & 167 & 89\% & 167 & 84\% & & &  &    \\ \botrule
\end{tabular}}
\end{table}

\section{Conclusions} 

Global statistical models of atomic masses, beta-decay lifetimes,
and nuclear spins and parities have been constructed using the
methodology of Support Vector Machines.  The predictive power
of these ``theory-thin'' models, which in essence are derived
from the data and only the data, is shown to be competitive
with, or superior to, that of conventional ``theory-thick'' models 
based on nuclear theory and phenomenology.  Conservative many-body
theorists may be troubled by the ``black-box'' nature of the SVM 
predictors, i.e., the impenetrability of their computational
machinery.  However, this alternative, highly pragmatic approach may 
represent a wave of the future in many fields of science -- already 
visible in the proliferation of density-functional computational packages 
for materials physics and eventually molecular biology, which,
for the user, are effectively black boxes.  While it is true
that the statistical models produced by advances in machine 
learning do not as yet yield the physical insights of traditional 
modeling approaches, their prospects for revealing new regularities 
of nature are by no means sterile.

\section*{Acknowledgements}

This research has received support from the U.S. National Science
Foundation under Grant No. PHY-0140316.  We acknowledge helpful
discussions and communications with S. Athanassopoulos, M. Binder,
E. Mavrommatis, T. Papenbrock, S. C. Pieper, and R. B. Wiringa.
In the regression studies reported herein, we have found the mySVM 
software and instruction manual created by S. R\"uping\cite{rueping} 
(Dortmund) to be very useful.

{}
\end{document}